\documentclass[12pt]{article}
\usepackage{graphicx}
\usepackage[superscript]{cite}

\newcommand{\dir}{Figs}

\newcommand{\ud}{\mbox{d}}

\newcommand{\vu}{\mbox{$\mathbf{u}$}}
\newcommand{\vr}{\mbox{$\mathbf{r}$}}
\newcommand{\vq}{\mbox{$\mathbf{q}$}}
\newcommand{\vn}{\mbox{$\mathbf{n}$}}
\newcommand{\vN}{\mbox{$\mathbf{N}$}}
\newcommand{\hq}{\mbox{$\hat{\mathbf{q}}$}}
\newcommand{\Tr}{\mbox{Tr}}
\newcommand{\QQ}{\mbox{${\bf Q}$}}
\newcommand{\kB}{\mbox{${k_{_B}}$}}

\begin{document}

\hspace*{1.5cm}
\begin{minipage}{13cm}
{\Large \bf Fluctuating Interfaces in Liquid Crystals}

\vspace{2\baselineskip}

{\em Friederike Schmid,${}^{*1}$ Guido Germano,${}^{1,2}$ Stefan
  Wolfsheimer,${}^{3,4}$ Tanja Schilling${}^{3}$}

\vspace{\baselineskip} 

${}^1$Physics Department, University of Bielefeld, Universit\"atsstrasse 25, 
D-33615 Bielefeld, Germany \\
Fax: (+49)521 1066455; E-mail: schmid@physik.uni-bielefeld.de\\
${}^2$Department of Chemistry, Philipps-Universit\"at Marburg, D-35032
Marburg, Germany \\
${}^3$Institute for Physics, Johannes-Gutenberg Universit\"at, Staudinger Weg 7,
D-55088 Mainz, Germany \\
${}^4$ Institute for Theoretical Physics,
Universit\"at G\"ottingen,
Friedrich-Hund-Platz 1,
37077 G\"ottingen, Germany

\vspace{\baselineskip}

{\bf Summary:}
We review and compare recent work on the properties of fluctuating interfaces between nematic 
and isotropic liquid-crystalline phases. Molecular dynamics and Monte Carlo simulations
have been carried out for systems of ellipsoids and hard rods with aspect ratio 15:1,
and the fluctuation spectrum of interface positions (the capillary wave spectrum) has 
been analyzed. In addition, the capillary wave spectrum has been calculated analytically
within the Landau-de Gennes theory. The theory predicts that the interfacial fluctuations
can be described in terms of a wave vector dependent interfacial tension, which is
anisotropic at small wavelengths (stiff director regime) and becomes isotropic at large 
wavelengths (flexible director regime). After determining the elastic constants in the
nematic phase, theory and simulation can be compared quantitatively. We obtain good
agreement for the stiff director regime. The crossover to the flexible director regime
is expected at wavelengths of the order of several thousand particle diameters, which
was not accessible to our simulations. 

\vspace{\baselineskip}

\end{minipage}

\vspace{1.5\baselineskip}


{\large \bf Introduction}

Elongated particles form liquid crystalline structures at high densities. For example, they often
exhibit a nematic phase, where the particles align along one preferred direction (the director), 
while their positions are disordered, like in a fluid. For symmetry reasons, the phase transition
between the nematic phase (N) and the fully disordered isotropic phase (I) must be first 
order\cite{degennes}. Hence the two phases coexist in a certain parameter regime, and are 
separated by a nematic-isotropic (NI)-interface. The properties of that interface are intriguing 
for several reasons: Since it separates two fluid phases, the interfacial tension of a planar 
interface of given orientation $\vN$ should be isotropic -- the free energy 
cost of increasing the interfacial area should not depend on the direction in which the area has 
been extended. On the other hand, the interface interacts with the adjacent, anisotropic, nematic 
fluid -- for example, it orients the director (``surface anchoring'') -- and one would expect 
this to have an influence on the interface. Hence, it should also exhibit anisotropic features.

The role of the interfacial tension for the interfacial structure can be assessed by studying 
the fluctuations of the interface positions, the ``capillary waves''. For interfaces between 
two simple fluids, the theory of capillary waves is quite simple: Let us consider a planar 
interface oriented in the $z$-direction with the projected area $A$, neglect overhangs, and 
parametrize the local interface position by a single-valued function $h(x,y)$. 
Fluctuations enlarge the interfacial area and thus cost the free energy\cite{smoluchowsky,buff,weeks}
\begin{equation}
\label{eq:f_cw}
{\cal F}\{h\} = \gamma \int \!\! \ud x\: \ud y\:
\sqrt{1+ (\partial_x h)^2 + (\partial_y h)^2 }
\approx \gamma A + \frac{\gamma}{2} \int \!\! \ud x\: \ud y\:
\big[ (\partial_x h)^2 + (\partial_y h)^2 \big]
\end{equation}
with the interfacial tension $\gamma$. Based on this functional, the thermal height
fluctuations $\langle | h(\vq) |^2 \rangle$ at given wave vector $\vq$ can
be calculated in a straightforward manner. One obtains the capillary wave spectrum
\begin{equation}
\label{eq:hq2_simple}
\langle | h(\vq) |^2 \rangle = \kB T/\gamma q^2.
\end{equation}
At a NI interface, the situation is more complicated. The capillary wave fluctuations 
are then determined by at least three factors, namely, (i) the interfacial tension, 
(ii) the surface anchoring, and (iii) the elasticity of the nematic bulk,
i.e., the elastic response of the nematic fluid to local director variations.

In the present paper, we shall review recent work that has shed light on the interplay of 
these three factors. The paper is organized as follows. In the next section, we discuss
large-scale simulations of NI interfaces in model liquid crystals, which have allowed to 
study in detail the capillary wave spectrum in these systems. In section three, a Landau-de Gennes 
theory of capillary waves is presented. The predictions of this theory are compared with the 
simulations in section four. Finally, we summarize and conclude with a brief outlook on 
nonequilibrium interfaces.

\vspace*{\baselineskip}

{\large \bf Computer Simulations}

Computer simulations of equilibrium NI interfaces were carried out for two standard model 
liquid crystals: Systems of ellipsoids\cite{mcdonald,akino} and systems of 
spherocylinders\cite{vink1,vink2,wolfsheimer}. In two of these studies\cite{akino,wolfsheimer}, 
the system sizes were large enough that capillary waves could be investigated in detail. 
These shall be compared with each other.

The model ellipsoids interact with each other {\em via} a soft, repulsive, 
Weeks-Chandler-Andersen-type\cite{wca} potential
\begin{equation}
\label{eq:potential}
V(\vu_i,\vu_j,\vr_{ij})=
\left\{ \begin{array}{ll}
 4\epsilon (X^{-12}-X^{-6}+\frac{1}{4}) & \mbox { $X^{6}>\frac{1}{2}$}  \\
     0                                & \mbox { otherwise}
         \end{array}
\right.
\end{equation}
with
\begin{equation}
X=\frac{r-\sigma(\vu_i,\vu_j,\hat{\vr}_{ij})+\sigma_s}{\sigma_s},
\end{equation}
where $\sigma_s$ is the diameter of the ellipsoids, $\vu_i$ and $\vu_j$ are the orientations of 
ellipsoids $i$ and $j$, $\hat{\vr}_{ij}$ is the unit vector connecting the centers of the two 
ellipsoids, and the shape function\cite{berne}
\begin{equation}
\label{eq:sigma}
\sigma({\bf u}_{i},{\bf u}_{j},{\hat{\bf r}})  = 
\sigma_{s} \Big\{ 1 - \frac{\chi}{2}
                         \Big[ \frac{ ({\bf u}_{i} \cdot {\hat{\bf r}}
                                  + {\bf u}_{j} \cdot {\hat{\bf r}})^{2} }
                                 {1+\chi {\bf u}_{i}\cdot{\bf u}_{j}}
                            +  \frac{ ({\bf u}_{i} \cdot {\hat{\bf r}}
                                  - {\bf u}_{j} \cdot {\hat{\bf r}})^{2} }
                             {1-\chi {\bf u}_{i}\cdot{\bf u}_{j}}
                     \Big] \Big\} ^{-1/2}
\end{equation}
approximates the contact distance between hard ellipsoids of elongation $\kappa$ with
$\chi = (\kappa^2-1)/(\kappa^2+1)$. The temperature was chosen $\kB T = \epsilon$.

The spherocylinders are taken to be hard rods with semi-spherical caps, i.e., lines of length 
$L$, which may not come closer to each other than a distance $D$ (the diameter of the rods). 

The elongation of the particles, $\kappa$ or $L/D$, respectively, was chosen 15 in both cases.
The simulations were carried out in the $NVT$-ensemble with roughly $N \approx 100.000$
particles in an elongated simulation box with side length ratios $L_x:L_y:L_z \equiv 1:1:2$
and periodic boundary conditions. The simulation method was Molecular dynamics in the case of the 
ellipsoids\cite{akino}, and Monte Carlo in the case of the hard rods\cite{wolfsheimer}. By choosing 
a mean density between the densities of the nematic and isotropic phase at coexistence, phase 
separation into a nematic and an isotropic slab was enforced. The equilibrated systems thus 
contained two planar NI interfaces with orientation normals $\vN$ parallel to the long axis of 
the simulation box. Both for ellipsoids and hard rods, the director in the nematic phase 
spontaneously aligned in the direction parallel to the interface (planar anchoring). Normal 
anchoring could be studied as well, but had to be enforced, e.g., by choosing special boundary 
conditions\cite{mcdonald}, or by preparing an initial configuration that contains a nematic slab 
with normal orientation\cite{wolfsheimer}. Here, we will almost exclusively discuss planar anchoring.
To analyze the capillary wave spectrum, the simulation box was split into columns (blocks) 
$B \times B \times L_z$, and the local interface positions were determined separately in each block. 
This gave two interface topographies $h(x,y)$ for every configuration, which could then be Fourier 
transformed to obtain the capillary wave spectrum, $|h(\vq)|^2$ (cf. Appendix).
More details on the simulation and the data analysis can be found in Refs.~\citeonline{akino} 
and \citeonline{wolfsheimer}.

Table \ref{tab:NI} summarizes the main properties of our model systems at coexistence.
Most of the data are compiled from earlier publications\cite{mcdonald,akino,vink2,wolfsheimer}, 
but the table also shows new, previously unpublished results. In particular, it gives
the Frank elastic constants\cite{degennes,frank} $K_1,K_2,K_3$ of the ellipsoid system in the 
nematic phase, which we have evaluated in order to use them for a quantitative comparison between 
theory and simulation (see below). To this end, separate simulations of a homogeneous system at 
the density of the nematic phase were conducted, and the order tensor fluctuations were analyzed, 
following a procedure described in Refs.~\citeonline{allen} and \citeonline{phuong}.
\begin{table}[h]
\caption{\label{tab:NI} Properties of our model systems at NI coexistence.
Numbers in brackets indicate errors in the last digit.}

\begin{tabular}[h]{c|c|c}
\hline
 & Ellipsoids & Hard rods \\
\hline 
densities & $\rho_N = 0.0181(1)/\sigma_s^3$ & $\rho_N = 0.027/D^3$  \\
& $\rho_I = 0.0149(1)/\sigma_s^3$ & $\rho_I = 0.023/D^3$ \\
\hline
reduced densities& $\rho_N^* = 0.192$ & $\rho_N^* = 0.22$  \\
& $\rho_I^* = 0.158$ & $\rho_I^*=0.193$ \\
$\rho^*=\rho/\rho_{cp}$ &
$\rho_{cp} =  \sqrt{2}/15$ &
$\rho_{cp} =  2/(\sqrt{2}+15 \sqrt{3})$ \\
\hline
order parameter & $S_N = 0.74$\cite{mcdonald} & $S_N \approx 0.7$ \\
\hline
Frank & $K_1 = 0.78(2)$ \hspace*{2.1cm} & \\
elastic constants & $K_2 = 0.32(2)$  at  $\rho = 0.018$ & \\
(nematic phase) & $K_3 = 3.61(5)$ \hspace*{2.1cm} & \\
\hline
interfacial tension & from pressure tensor\cite{mcdonald,akino} 
& from histogram method\cite{vink2} \\
\cline{2-3}
planar anchoring${}^\dagger$ 
& $\gamma_{IN} = 0.011(4) \kB T/\sigma_s^2$\cite{mcdonald} & 
          $\gamma_{IN} = 0.0064 \kB T/D^2$\cite{vink2}
 \\
       & $\gamma_{IN} = 0.009(3) \kB T/\sigma_s^2$\cite{akino} 
&           \\
\cline{2-3}
normal anchoring & $\gamma_{IN} = 0.015(4) \kB T/\sigma_s^2$\cite{mcdonald}
& \\
\hline
interfacial width${}^\ddagger$
& $w \approx 7.3 \sigma_s$ & $w \approx 10.6 D$ \\
\end{tabular} \\
{\footnotesize
${}^\dagger$ The two values for the ellipsoids were obtained in simulations with different 
system sizes. They agree within the error.\\
${}^\ddagger$ The interfacial width has been determined by a fit of the local order parameter 
profile to a $\tanh$ profile, $\tanh( (z-z_0)/w$. Due to the effect of capillary waves, 
it depends on the lateral system size. The values given here correspond to local profiles 
evaluated in blocks of lateral size $B=2 \kappa \sigma_s$ (ellipsoids)\cite{akino} 
and $B=2 L$ (rods)\cite{wolfsheimer}.}
\end{table}

\begin{figure}[t]
\includegraphics[width=0.48\textwidth]{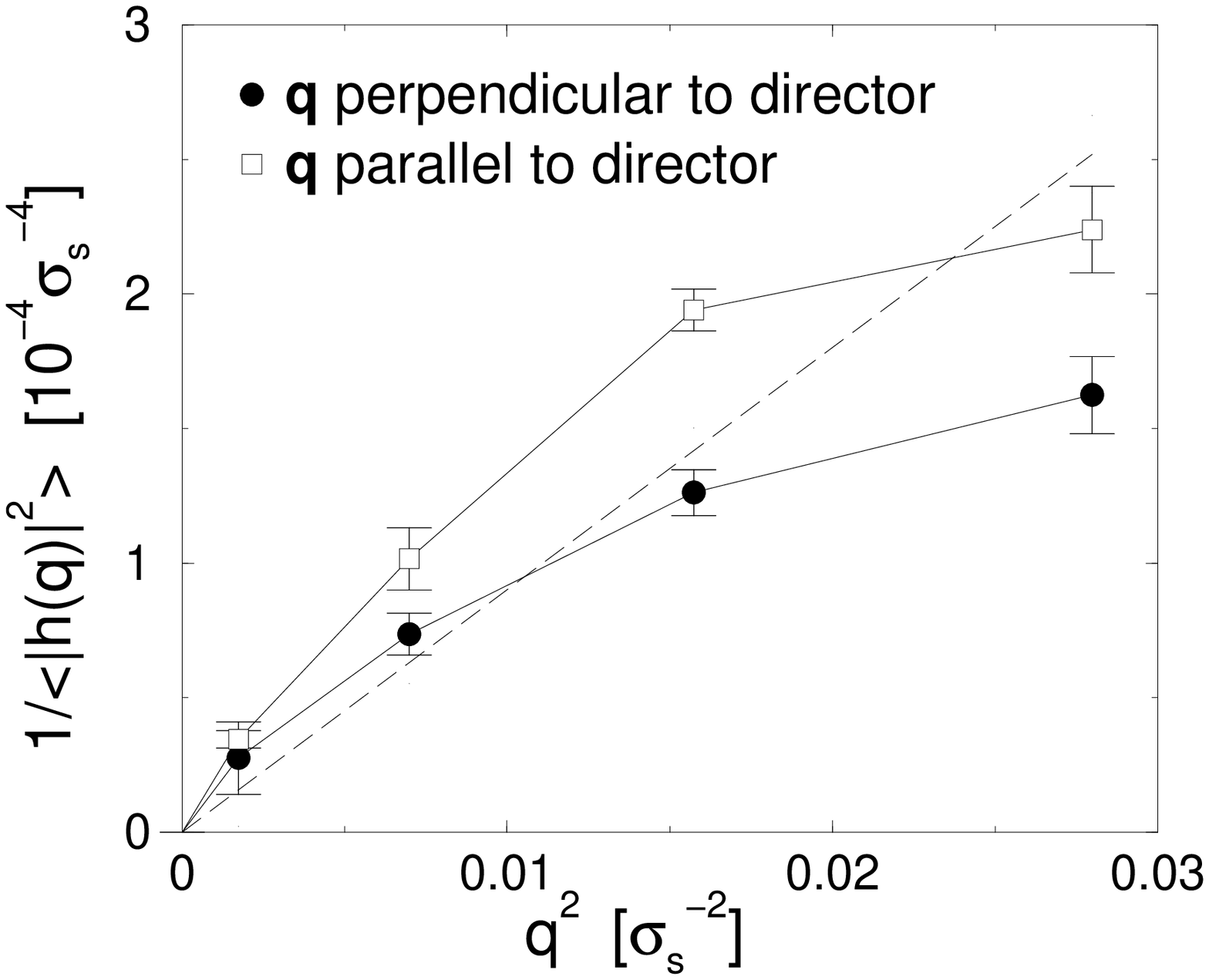} 
\hfill
\includegraphics[width=0.46\textwidth]{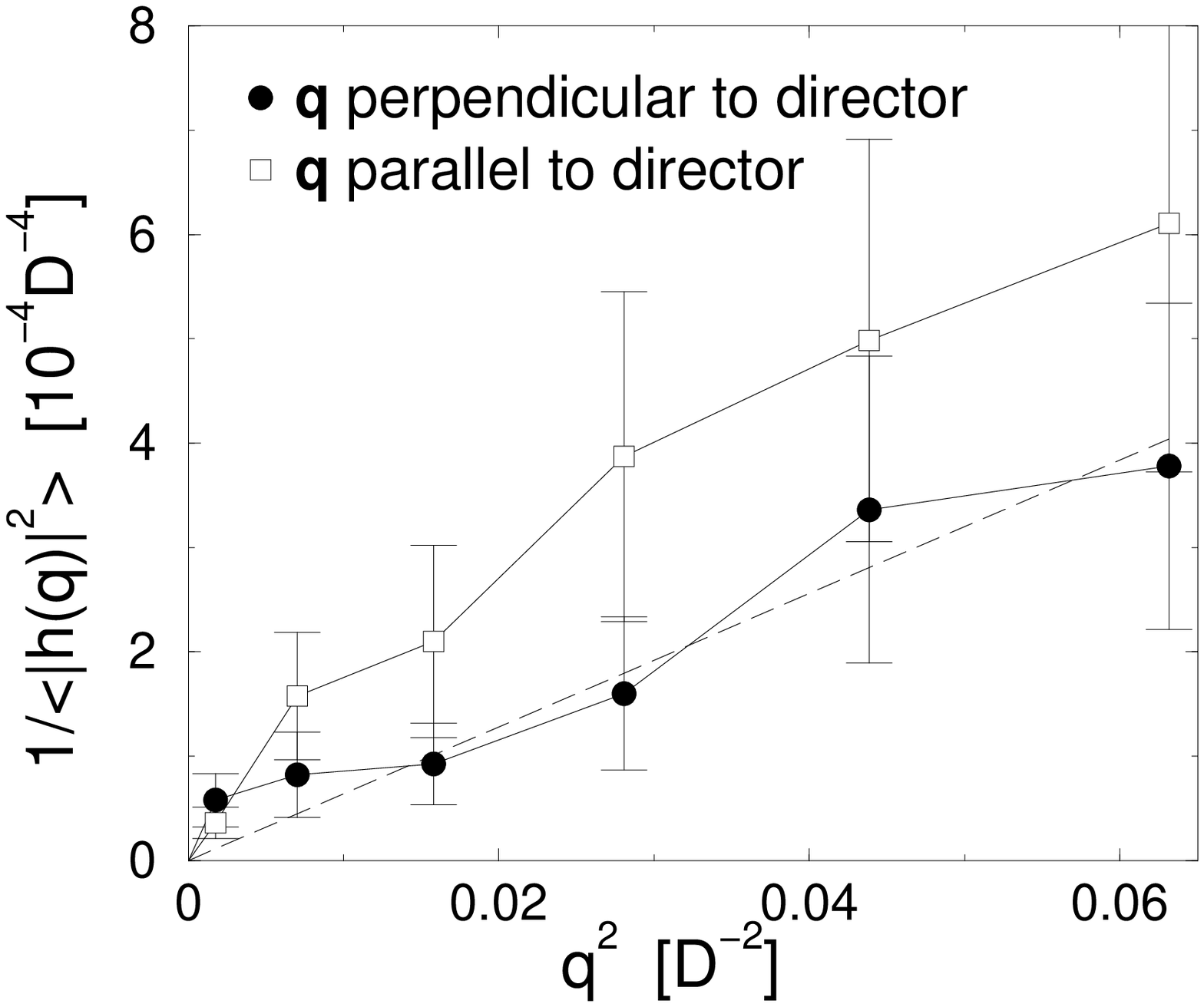} \\
\caption{\label{fig:cap_sim} Inverse mean-squared Fourier components of the interface position $h$
vs. wave vector \vq\ squared for \vq\ parallel and perpendicular to the director $\vn_0$.
Left: Ellipsoids (similar to Ref.~\citeonline{akino}; data taken from Ref.~\citeonline{germano}). 
Right: Hard rods (similar to Ref.~\citeonline{wolfsheimer}, different data set).
The straight dashed lines indicate the prediction of Eq.~(\ref{eq:hq2_simple}) with the
interfacial tension taken from Table~\ref{tab:NI}, $\gamma = 0.009 \kB T/\sigma_s^2$ (ellipsoids) 
and $\gamma = 0.0064 \kB T/D^2$ (hard rods). Remarkably, it seems to fit the lower branch 
($\vq \perp \vn_0$) rather well in both systems. 
}
\end{figure}

The capillary wave spectra for the two model systems are shown in Fig.~\ref{fig:cap_sim}. 
The capillary waves are clearly anisotropic: In the direction of the director, they
are suppressed by a factor 0.3-0.5, compared to those in the perpendicular direction. 
Nevertheless, they still roughly follow the proportionality law 
$1/\langle | h(\vq)|^2 \rangle \sim q^2$ predicted by Eq.~(\ref{eq:hq2_simple}).
Deviations are observed in the ellipsoid system for wavelengths 
comparable to the particle length, $2 \pi/q \sim 3 \kappa \sigma_s$. In the hard rod system, the 
proportionality persists over the whole accessible wave vector range. Hence it seems that 
Eq.~(\ref{eq:hq2_simple}) would provide a reasonable description of our data, provided the 
interfacial tension $\gamma$ were allowed to depend on the direction of the wave vector, \vq. 
On the other hand, we have pointed out in the introduction that the {\em macroscopic} surface 
tension between two fluid phases must be isotropic. To resolve this apparent contradiction, 
a Landau-de Gennes type theory of capillary wave fluctuations at NI interfaces has been
developed\cite{elgeti}. It shall be presented next.

\vspace*{\baselineskip}

{\large \bf Theory}

The Landau-de Gennes theory is based on a free energy expansion in powers of a symmetric and 
traceless (3$\times$3) order tensor field ${\bf Q}(\vr)$\cite{degennes} 
\begin{equation}
F = \int \!\!\! d^3\!\!r \, \Big\{ \frac{A}{2} \Tr(\QQ^2)
\! + \! \frac{B}{3} \Tr(\QQ^3)
\! + \! \frac{C}{4} \Tr(\QQ^2)^2
+\: \frac{L_1}{2} \partial_i Q_{jk} \partial_i Q_{jk}
+ \frac{L_2}{2} \partial_i Q_{ij} \partial_k Q_{kj} \Big\},
\label{eq:landau_q}
\end{equation}
with $B < 0$ close to the NI transition. If biaxiality can be neglected (both in our ellipsoid and 
hard rod system, the maximum value of the biaxiality near the NI interface was less than 0.06), the 
order tensor can be written as\cite{sheng} 
$Q_{ij}(\vr) = \frac{1}{2} S(\vr) (3 n_i(\vr) n_j(\vr) - \delta_{ij})$, where $S(\vr)$ is the local 
scalar order parameter, and \vn\ a unit vector characterizing the local director. For further
simplification, we introduce ``natural'' units 
$S_0 = - {2 B}/{9 C}$, $\xi_0 = 2\sqrt{{(L_1 + L_2/6)}/{3C}} \: {S_0^{-1}}$,
and $\epsilon_0 = ({3C}/{16}) S_0^4 \xi_0^3$ for the order parameter,
the length, and the energy, and rescale all quantities by these units.
The free energy functional (\ref{eq:landau_q}) then takes the form\cite{cheung}
\begin{eqnarray}
\label{eq:landau}
\lefteqn{
F = 3 \int d^3r \: \{ f + g_1 + g_2 \} \quad \mbox{with} \quad
f = S^2 ((S - 1)^2 + t), } \qquad 
\\ \nonumber
g_1 &=& (\nabla \cdot S)^2 + \alpha (\vn \cdot \nabla S)^2 +
          4 \alpha S \Big( (\nabla \cdot \vn) (\vn \cdot \nabla S)
          + \frac{1}{2} (\vn \times \nabla \times \vn) (\nabla S) \Big),
\\ \nonumber
g_2 &=& S^2 \Big(
(3 + 2 \alpha) (\nabla \vn)^2
+ (3 - \alpha) (\vn \cdot \nabla \times \vn)^2
+ \: (3 + 2 \alpha)(\vn \times \nabla \times \vn)^2
\Big).
\end{eqnarray}
with the two dimensionless parameters,
\begin{equation}
\label{eq:t-alpha}
t = \frac{1}{4} A  \frac{S_0^2 \xi_0^3}{\epsilon_0} - 1,
\qquad \mbox{and} \qquad
\alpha = \frac{1}{2} \: \frac{L_2}{(L_1 + L_2/6)}.
\end{equation}
The parameter $t$ gives the distance from NI coexistence and becomes zero at coexistence.
The parameter $\alpha$ characterizes the response of the system to order parameter and director
variations. In Eq.~(\ref{eq:landau}), the first term, $f(S)$, is the free energy density of a 
homogeneous system, the second term, $g_1$, accounts for the effect of order parameter variations, 
and the last term, $g_2$, corresponds to the Frank elastic energy of a nematic fluid with spatially 
varying director. This last term relates the free energy functional (\ref{eq:landau}) to the 
well-known Frank elastic energy\cite{degennes,frank} 
\begin{equation}
\label{eq:frank}
F_F = \frac{1}{2} \int \! d^3\!r \,
\Big\{ K_1 (\nabla \vn)^2 + K_2 (\vn\cdot\nabla\times\vn)^2 + K_3(\vn\times\nabla\times \vn)^2 \Big\}
\end{equation}
and allows to identify the three Frank elastic constants splay ($K_1$), twist ($K_2$), and bend
($K_3$). We note that Eq.~(\ref{eq:landau}) predicts $K_1=K_3$, whereas experimentally and
in simulations (see Table \ref{tab:NI}), the parameter $K_3$ is usually much higher than $K_1$.
To improve the theory in this respect, one would have to include higher powers of ${\bf Q}$ in
the expansion (\ref{eq:landau_q}).

Minimizing the free energy (\ref{eq:landau}) with the boundary conditions $S=0$ at $z \to -\infty$,
$S=1$ at $z \to \infty$ and fixed director $\vn \equiv \vn_0$ yields the mean-field structure 
of a planar NI interface at fixed anchoring angle $\cos \theta = \vn_{0,z} \equiv (\vn_0 \vN)$ 
($\vN$ being the interface normal): The order parameter adopts a tanh profile, 
$S(z) = S_0 \bar{S}(z/w)$ with
\begin{equation}
\label{eq:sbar}
\bar{S}(\tau) = \frac{1}{2}(1 + \tanh(\tau)) 
\qquad \mbox{and the width} \qquad
w=2 \xi_0 \sqrt{1 + \alpha (\vn_0 \vN)^2}. 
\end{equation}
The mean-field interfacial tension is 
\begin{equation}
\label{eq:tension}
\sigma = \sigma_0 \sqrt{1 + \alpha (\vn_0 \vN)^2} \qquad \mbox{with} \qquad
\sigma_0 = \epsilon_0/\xi_0^2. 
\end{equation}
The parameter $\alpha$ thus not only determines the elastic
constants, but also the strength and the direction of the anchoring at the interface. 
At $\alpha > 0$, the interface favors planar alignment, and at $\alpha < 0$, it favors normal 
alignment. 

To study capillary waves, we must allow the interfacial position to vary. We first consider a
simplified variant, where the director is still taken to be constant throughout the system, 
$\vn \equiv \vn_0$, and lies in the $(x,y)$ plane ($n_{0,z}=0$).
For the order parameter, we make the Ansatz $S(\vr) = S_0 \bar{S}\big[(z-h(x,y))/w\big]$, 
where $\bar{S}$ and $w$ are given by Eq. (\ref{eq:sbar}). We note that the local surface normal 
$\vN$ is no longer constant, but depends on the gradient of the local interface position $h(x,y)$, 
i.e., $\vN \propto (- \partial_x h,- \partial_y h,1)$. After inserting this Ansatz into the 
free energy functional (\ref{eq:landau}), Fourier transforming $(x,y) \to \vq$, and
defining $\hq = \vq/q$, we obtain
\begin{equation}
\label{eq:dir_const}
F = \sigma_0 A + \frac{1}{2} \int \!\! d^2 q \: |h(\vq)|^2 \: q^2 \: \gamma(\vq)
\quad \mbox{with} \quad
\gamma(\vq) := \sigma_0(1 + \alpha (\hq \vn_0)^2) ).
\end{equation}
The capillary wave spectrum can be calculated in complete analogy to the simple fluid
case\cite{elgeti}, Eq.~(\ref{eq:hq2_simple}), and one gets
\begin{equation}
\label{eq:hq2}
\langle | h(\vq) |^2 \rangle = \kB T/\gamma(\vq) \, q^2.
\end{equation}
Hence this simple approximation already produces an anisotropic capillary wave spectrum. It
predicts the proportionality $1/\langle |h(\vq)|^2 \rangle \propto q^2$ suggested by our
simulation data. Two remarks are in order here. First, the analogy to simple fluids is perfect, 
if the function $\gamma(\vq)$ is interpreted as a wave vector dependent interfacial tension. 
Second, $\gamma(\vq)$ only depends on the {\em orientation} of $\vq$, not on its absolute value. 
As a consequence, the capillary wave spectrum is scale invariant, all length scales are 
equivalent, and the interfacial tension is predicted to be anisotropic on all length 
scales. However, we have argued above that this is unphysical -- the macroscopic interfacial
tension must be isotropic. It turns out that the problem is caused by the constant director 
constraint. To resolve it, we must extend the theory such that the director is allowed to follow the 
interfacial undulations.

To this end, a second approximation has been adopted: The ``local profile'' approximation. 
The main assumption here is that the width of the interface is small, compared to the relevant
length scales of the capillary waves. This is of course questionable, since we have
just seen that capillary waves in simple systems are scale invariant. However, computer
simulations of various systems\cite{schmid,werner,mueller} have shown that the concept of
separating ``intrinsic profiles'' and capillary waves often provides a highly satisfactory 
{\em quantitative} description of interfacial structures. We separate the free energy 
(\ref{eq:landau}) into an interface and a bulk contribution, $F=F_I + F_B$. The bulk contribution,
\begin{equation}
\label{eq:fbulk}
F_B = \int_{h(x,y)}^{\infty} \!\!\! dz \int \! dx \: dy \:
\Big\{
(3 + 2 \alpha) (\nabla \vn)^2
+ (3 - \alpha) (\vn \cdot \nabla \times \vn)^2
+ \: (3 + 2 \alpha)(\vn \times \nabla \times \vn)^2
\Big\},
\end{equation}
accounts for the elastic energy in the nematic fluid. The remaining 
interface free energy $F_I$ vanishes far from the surface. It is evaluated using the assumption 
that the local order parameter profile has mean-field shape in the direction $\vN$ perpendicular 
to the interface, and that the director variations are slow, compared to the order parameter
variations in the vicinity of the interface, such that they can be approximated 
by a linear behavior in the interfacial region. The total free energy is then minimized with
respect to the director field $\vn(\vr)$ for fixed interfacial position $h(x,y)$,
and for given bulk director $\vn_0 = \lim_{z\to \infty}\vn(\vr)$. Since globally, we still have 
planar anchoring at $\alpha >0$, the bulk director $\vn_0$ lies in the $(x,y)$-plane.

Details on the calculation can be found in Ref.~\citeonline{elgeti}. Here we just sketch 
the main results: The final free energy as a function of $h(x,y)$ can be cast in the same
way as Eq.~(\ref{eq:dir_const}). However, the wave vector dependent interfacial tension
$\gamma(\vq)$ now depends on the absolute value of $q$. Expanded in powers of $q$, it takes
the form
\begin{equation}
\label{eq:ci}
\gamma(\vq) \approx \sigma_0 \Big( 1 + q \: C_3[(\hq \vn_0)^2] 
+ q^2 \: C_4[(\hq \vn_0)^2] + \cdots \Big),
\end{equation}
The leading term is isotropic. The capillary wave spectrum, still given by Eq.~(\ref{eq:hq2}), 
remains anisotropic, but the anisotropy comes in through the higher order contributions to 
$\gamma(\vq)$. Thus the improved theory resolves our problem: At large wavelengths, the 
interfacial tension becomes isotropic and assumes the value $\gamma = \sigma_0$. 
This is also the surface free energy per area that one would obtain in a macroscopic 
measurement~\cite{footnote}. It is worth noting that in the direction {\em perpendicular} 
to the director ($\hq \perp \vn_0$), the $C_i$ in Eq.~(\ref{eq:ci}) vanish\cite{elgeti}, such
that $\gamma(\vq)$ is equal to $\sigma_0$ at {\em all} wavelengths. 
In that direction, the capillary wave spectrum corresponds to that of a simple interface, 
which is solely determined by the macroscopic surface tension $\gamma$ (Eq.~\ref{eq:hq2_simple}).

These results are gratifying. However, the predicted capillary wave spectrum is now in
apparent disagreement with the simulation data, which seem to point towards a general
$\langle |h(\vq)|^2 \rangle \propto 1/q^2$ dependence. To assess the apparent discrepancy, we
must compare the theory and the simulation data at a quantitative level.

\vspace*{\baselineskip}

{\large \bf Comparison between Theory and Simulation}

At coexistence, the theory has only one dimensionless parameter, $\alpha$. Since it enters the 
ratios of the elastic constants as well as the interfacial anchoring parameters, we have several 
independent ways of determining its numerical value. In the following, we shall estimate $\alpha$ for
the ellipsoid system, based on the data collected in Table \ref{tab:NI}. For the hard rod
system, the elastic constants and the anchoring parameters are not yet available -- for
particles as elongated as ours, however, the numerical value for $\alpha$ should be comparable.

According to the data of Ref.~\citeonline{mcdonald}, the ratio of interfacial tensions 
for normal and planar anchoring is roughly given by 
$\gamma_{NI}^{\mbox{\tiny normal}}/\gamma_{NI}^{\mbox{\tiny planar}} \approx 1.4$. 
Comparing this with 
$\gamma_{NI}^{\mbox{\tiny normal}}/\gamma_{NI}^{\mbox{\tiny planar}} = \sqrt{1+\alpha}$,
we get $\alpha \sim 1$.
Two other estimates for $\alpha$ are provided by the numerical values of the elastic constants.
According to the theory, the ratio of $K_1$ or $K_3$ and $K_2$ is 
$K_{1,3}/K_2 = (3+2 \alpha)/(3-\alpha)$. Numerically, $K_1$ and $K_3$ are different, hence 
we obtain two different values for $\alpha$. From $K_3/K_2 \sim 11.3$, we get $\alpha \sim 2.3$, 
and from $K_1/K_2 \sim 2.43$, we get $\alpha \sim 1$. Comparing the three estimates,
we conclude that the choice $\alpha = 1.$ is reasonable. This can now be inserted in the
theory.

\begin{figure}[t]
\includegraphics[width=0.49\textwidth]{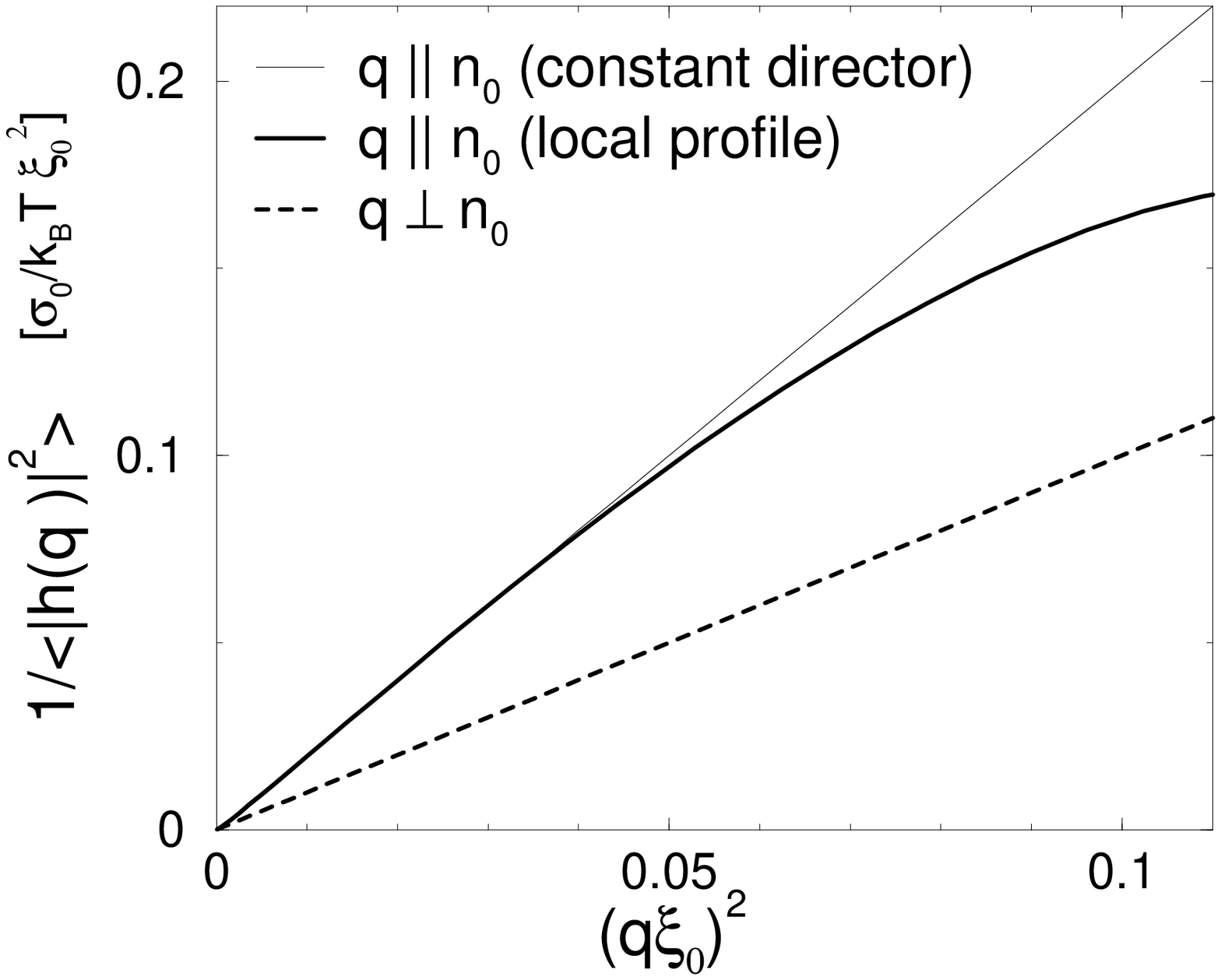} 
\hfill
\includegraphics[width=0.49\textwidth]{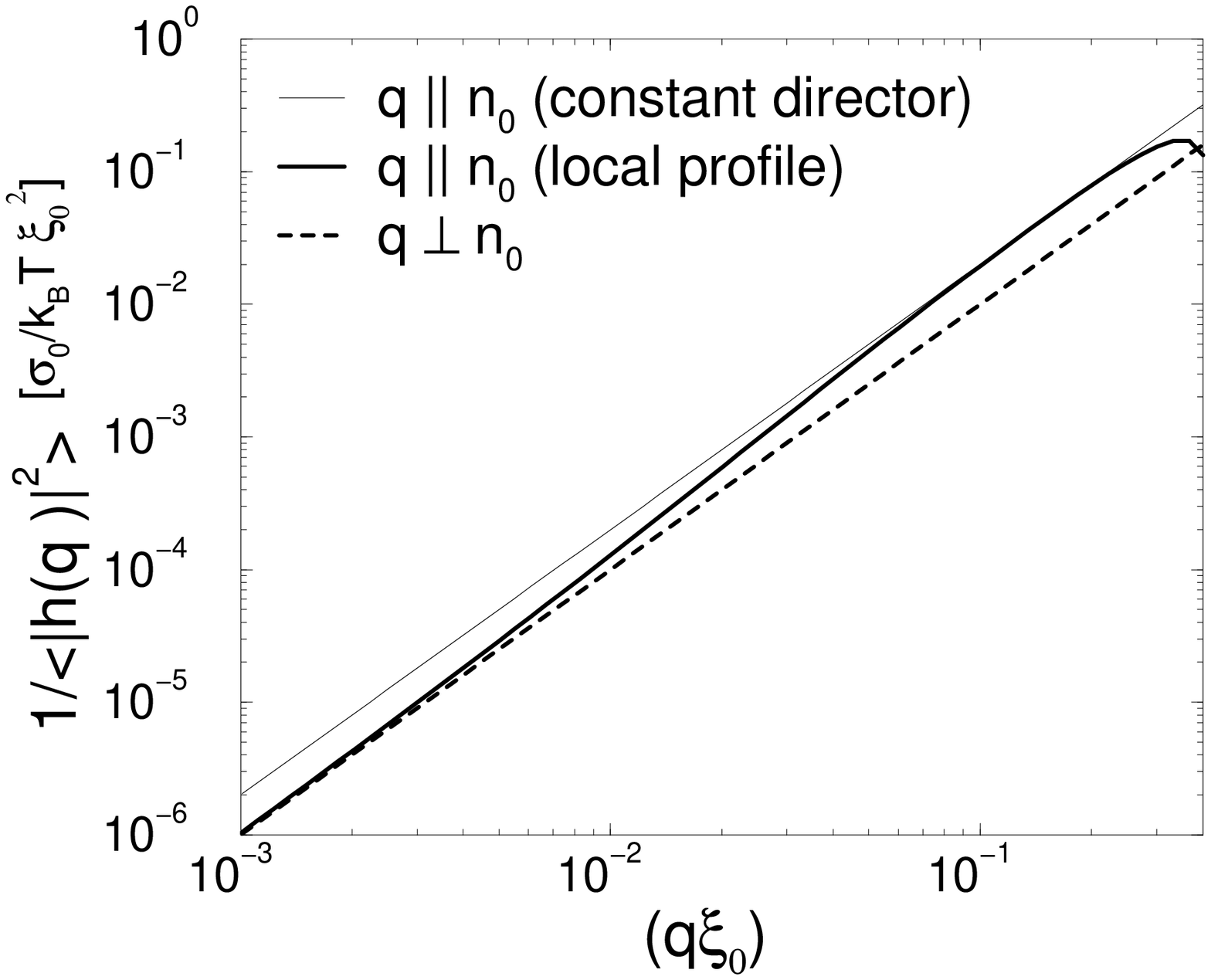} \\
\caption{\label{fig:cap_theory} Capillary wave spectrum in the direction parallel and perpendicular 
to the director as predicted by the theory for $\alpha = 1.$ The thin solid line corresponds to 
the constant director approximation, Eqs.~(\ref{eq:dir_const}) and (\ref{eq:hq2}), the thick solid 
line to the local profile approximation. In the direction perpendicular to the director, both 
approximations give the same result, which also agrees with the prediction for simple fluid
interfaces, Eq.~(\ref{eq:hq2_simple}). Left: Linear plot vs.\ $q^2$. 
Right: Double logarithmic plot showing the crossover between isotropic and anisotropic behavior.
}
\end{figure}

Fig.~\ref{fig:cap_theory} shows the corresponding theoretical curves for the capillary wave 
spectrum. As discussed above, the spectrum becomes isotropic in the limit
of large wavelengths, $q \to 0$. However, the crossover occurs in the $q$-range 
$q \xi_0 \sim 10^{-2}$, corresponding to the length scale $2 \pi/q \sim 600 \xi_0$. 
Isotropic behavior is expected for wave vectors less than $q \xi_0 \sim 3 \cdot 10^{-3}$, i.e.,
length scales larger than $\sim 2000 \xi_0$. Taking into account that $\xi_0$, is roughly half 
the interfacial width, the theory would thus predict an isotropic capillary wave spectrum 
on the length scale of $\sim 7000$ particle diameters in the ellipsoid system, or 
$\sim 10000$ particle diameters in the hard rod system. Hence it is not surprising, that this 
regime has not been observed in the simulations. In the $q$-range $q \xi_0 \sim 0.05-0.2$ 
(corresponding to $2 \pi/q \stackrel{<}{\sim} 100 \xi_0$), the capillary waves predicted 
by the local profile theory follow closely those of the constant director approximation: The 
director is effectively stiff. The elastic penalty on director variations is sufficiently strong 
that it prevents the director from following the interfacial undulations. At large wave vectors
$q \xi_0 \sim 0.4$, the curve predicted by the local profile approximation drops sharply. This 
is an artefact, the approximation breaks down for such small length scales~\cite{elgeti}. We
recall that the local profile Ansatz is based on the assumption of separate ``interfacial`` and 
``capillary wave`` length scales, and is thus bound to fail as $q \xi_0$ approaches one.

Comparing the theoretical capillary wave spectrum, Fig.~\ref{fig:cap_theory}, with the simulation 
results, Fig.~\ref{fig:cap_sim}, and disregarding the high $q$-regime, we find reasonable agreement
between theory and simulations. The theory predicts that the capillary waves in the direction
of the director should be suppressed by a factor of the order two, which is roughly reproduced by
the simulations. It explains why the simulations fail to produce an isotropic fluctuation 
spectrum at large wavelengths, and why the fluctuations in the direction perpendicular to the 
director are rather well described by the simplest capillary wave theory for isotropic fluid 
interfaces, Eq.~(\ref{eq:hq2_simple}), if one inserts the global surface tension for $\gamma$. 
Moreover, it gives a reason why the capillary waves in the direction parallel to the director
may show a $1/q^2$ behavior even in the anisotropic regime, as has been observed in the hard
rod system. 

\vspace*{\baselineskip}

{\large \bf Discussion and Summary}

In the present paper, we have reviewed and reexamined recent work on interfacial fluctuations 
(capillary wave fluctuations) of NI interfaces in situations where the bulk director is on average 
oriented parallel to the interface. In the past, we have studied such interfaces by computer 
simulations of two model liquid crystals, and by an analytical continuum theory. Here, we
compare the results of the different studies and relate them to each other. The interfacial 
capillary wave fluctuations of NI interfaces are governed by the competition of the interfacial 
tension, the anchoring of the nematic director at the interface, and the elasticity of the director 
in the bulk. As a result, they are anisotropic, the fluctuations are suppressed in the direction 
parallel to the director.  If the director were infinitely stiff, the anisotropy would persist 
on all length scales. Relaxing the director relieves this behavior, and the fluctuations become 
isotropic for large wavelengths. Hence we can identify two regimes: A ``flexible director'' regime, 
where the capillary waves are isotropic and governed by the macroscopic surface tension, and
a ``stiff director'' regime, where they are anisotropic and governed by a ``surface tension''
$\gamma(\vq)$ which depends on the orientation and the magnitude of the wave vector.
The crossover length scale between the two regimes is very large, such that it could not 
bee observed in the simulations. Otherwise, the simulations are in good semiquantitative 
agreement with the theory.

We thank Nobuhiko Akino, Michael Allen, Andrew McDonald, Jens Elgeti, and Richard Vink for enjoyable and fruitful collaborations that have led to the results presented
in this paper. This work was funded by the Deutsche Forschungsgemeinschaft (DFG). 
The simulations were carried out at the John-von-Neumann computing center at the
Forschungszentrum J\"ulich.

\vspace*{\baselineskip}

{\large \bf Appendix: Fourier transform of $h(x,y)$}

In this appendix we shall briefly sketch how the data for the interfacial height fluctuations
$h(x,y)$ have to be Fourier transformed in order to ensure the validity of the simple expressions 
(\ref{eq:hq2_simple}) and (\ref{eq:hq2}). Given a planar interface with the projected
area $A = L_x L_y$, and a discrete data set $h(x_n,y_m)$ with points $(x_n,y_m)$
on a regular grid ($n \in [1,N], m \in [1,M], x_n = L_x n/N, y_m = L_y m/M$),
then the Fourier transform $h(q_k,q_l)$ is defined by
\begin{equation}
\label{eq:fourier}
h(q_k,q_l) = \frac{\sqrt{A}}{N M} \sum_{n,m} \mbox{e}^{i (q_k x_n + q_l y_m)} h(x_n,y_m)
\end{equation}
with $q$-values $q_k = 2 \pi k/L_x$, $q_l = 2 \pi l/L_y$ and $k \in [1,N], l \in [1,M]$. 
The inverse Fourier transform is
$
h(x_n,y_m) = \frac{1}{\sqrt{A}} \sum_{k,l} \mbox{e}^{-i (q_k x_n + q_l y_m)} h(q_k,q_l).
$
We note that $h(q_k,q_l)$ has the unit of a {\em squared} length. The convention
(\ref{eq:fourier}) differs from the symmetrical definition
$h(q_k,q_l) = \frac{1}{\sqrt{N M}} \sum_{n,m} \mbox{e}^{i (q_k x_n + q_l y_m)} h(x_n,y_m)$,
which is often used for discrete Fourier transforms. When choosing the latter, one
has to introduce an additional prefactor $A/NM$ in Eqs.~(\ref{eq:hq2_simple}) and (\ref{eq:hq2}).

\vspace*{-\baselineskip}

\renewcommand{\refname}{}


\begin{thebibliography}{99}
\bibitem{footnote} This is actually a self-consistency requirement on the microscopic
``coarse-graining'' length of the capillary wave theory.
\end{thebibliography}
\end{document}